

\input phyzzx

\def\np{Nucl. Phys.}
\def\pl{Phys. Lett.}

\def\pr{Phys. Rev.}
\def\ap{Ann. Phys.}
\def\cmp{Comm. Math. Phys.}
\def\ijmp{Int. J. Mod. Phys.}
\def\mpl{Mod. Phys. Lett.}

\def\ex{{\hbox{\rm e}}}

\def\tr{{\hbox{\rm Tr}}}

\def\to{{\rightarrow}}

\tolerance=500000
\overfullrule=0pt

\Pubnum={US-FT-10/91}
\pubnum={US-FT-10/91}
\date={October, 1991}
\pubtype={}
\titlepage

\title{ANALYSIS OF OBSERVABLES IN CHERN-SIMONS PERTURBATION THEORY}
\author{ M. Alvarez and J.M.F.
Labastida\foot{E-mail: LABASTIDA@EUSCVX.DECNET} } \address{Departamento de
F\'\i sica de Part\'\i culas\break Universidade de Santiago\break E-15706
Santiago de Compostela, Spain}

\abstract{Chern-Simons Theory with gauge group $SU(N)$
is analyzed from a perturbation theory point of view. The vacuum
expectation value of the unknot is computed up to order $g^6$ and it is
shown that agreement with the exact result by Witten implies no
quantum correction at two loops for the two-point function. In addition,
it is shown from a perturbation theory point of view that the framing
dependence of the vacuum expectation value of an arbitrary knot factorizes
in the form predicted by Witten.}

\endpage
\pagenumber=1

\chapter{Introduction}

Chern-Simons gauge theory was solved exactly by Witten
\REF\witCS{E. Witten \journal\cmp&121(89)351}
[\witCS] using non-perturbative methods. This solution has been obtained
subsequently by other groups using both, the point of view of canonical
quantization \REF\bos{M. Bos and V.P. Nair \journal\pl&B223(89)61
\journal\ijmp&A5(90)959}
\REF\mur{H. Murayama, ``Explicit Quantization of the Chern-Simons
Action", University of Tokyo preprint, UT-542, March 1989}
\REF\lr{J.M.F. Labastida and A.V. Ramallo \journal\pl&B227(89)92
\journal\pl&B228(89)214}
\REF\emss{S. Elitzur, G. Moore, A. Schwimmer and N. Seiberg
\journal\np&B326(89)108}
\REF\fk{J. Frohlich and C. King \journal\cmp&126(89)167}
\REF\djt{G.V. Dunne, R. Jackiw, and
C.A. Trugenberger \journal\ap&194(89)197}
\REF\pietra{S. Axelrod, S. Della Pietra and E. Witten,
{\sl J. Diff. Geom.} {\bf 33} (1991) 787}
\REF\llr{J.M.F. Labastida, P.M. Llatas and A.V. Ramallo
\journal\np&B348(91)651}
\REF\gaku{K. Gawedzki and A. Kupiainen \journal\cmp&135(91)531}
[\bos-\gaku], and of
current algebra   \REF\mszoo{G. Moore and N. Seiberg
\journal\pl&B220(89)422} [\mszoo,\emss] as originally proposed in
[\witCS]. The exact result for the vacuum expectation value of the
observables of the theory is analytic in the inverse of the Chern-Simons
parameter $k$. Defining the Chern-Simons coupling constant as
$g=\sqrt{4\pi/k}$ the exact result  suggests that the
small coupling constant perturbation expansion should reproduce the exact
result. One does not expect any non-perturbative effect in Chern-Simons
gauge theory. Perturbative approaches to the theory under consideration
have been carried out during  the last two years
\REF\gmm{E. Guadagnini, M. Martellini
and M. Mintchev\journal\pl&B227(89)111 \journal\pl&B228(89)489
\journal\np&B330(90)575} \REF\alr{L. Alvarez-Gaum\'e, J.M.F. Labastida,
and A.V. Ramallo \journal\np&B334(90)103, {\sl \np} {B} (Proc. Suppl.)
{\bf 18B} (1990) 1}
\REF\prao{R.D. Prisarski and S. Rao\journal\pr&D22(85)208}
\REF\brt{D. Birmingham, M. Rakowski and G. Thompson
\journal\np&B234(90)83}
\REF\sor{F. Delduc, F. Gieres and S.P. Sorella\journal\pl&B225(89)367}
\REF\dr{P.H. Damgaard and V.O. Rivelles\journal\pl&B245(89)48}
\REF\cpm{C.P. Mart\'\i n
\journal\pl&B241(90)513}
\REF\aso{M. Asorey and F. Falceto
\journal\pl&B241(90)31}
\REF\bc{A. Blasi and R. Collina
\journal\np&B345(90)472}
\REF\dgs{F. Delduc, C. Lucchesi, O. Piguet and S.P. Sorella
\journal\np&B346(90)513}
\REF\dd{D. Daniel and N. Dorey\journal\pl&B246(90)82}
\REF\dor{N. Dorey\journal\pl&B246(90)87}
\REF\cpmdos{C.P. Mart\'\i n\journal\pl&B263(91)69}
\REF\dror{D. Bar-Natan, ``Perturbative Chern-Simons
Theory", Princeton preprint, 1990}
[\gmm-\dror]. The main question which have been
addressed in these works is which one is the renormalization scheme which
leads to the exact result obtained by Witten. In Chern-Simons gauge
theory there are two problems which must be taken into account.
On the one hand, the loop expansion possesses divergences already at one
loop for two-point functions which must be regularized. On the other hand, some
of the observables of the theory, the Wilson lines, possess products of
operators at  coincident points in their integration regions. The loop
expansion  divergences must be regulated in perturbation theory to obtain a
finite answer to be compared to the exact result. The
ambiguities present when considering products of operators at coincident points
forces  to make a choice in defining the observables of the theory. The main
goal of this paper is to give a regularization procedure and a choice to solve
the problem of the ambiguity when  considering products of operators at
coincident points whose perturbative expansion coincides with the exact result
[\witCS]. The second aspect of the problem was solved successfully in [\gmm]
and we will follow here their approach.

To handle the ambiguity associated to products of operators at
coincident points one must consider
framed links instead of links \REF\pol{A. M. Polyakov
\journal\mpl&A3(88)325} [\pol,\witCS,\gmm], in other words one must
introduce a band instead of a knot and the corresponding integer number $n$
which indicates the number of times that the band is twisted. In the
non-perturbative approach leading to the exact result [\witCS] the origin of
the dependence on the framing comes about because one must construct the
observables on the surface of a Riemann surface which then must be glued to
another Riemann surface to build a three-dimensional manifold. The same
knot can be obtained using that procedure in a variety of ways leading to
different quantities for  observables which, however,  differ by a factor
which  is associated to the framing. As first found out in [\witCS] this
factor is just $\exp(2\pi i n h)$ where $h$ is the conformal weight  of the
representation (for $SU(N)$ in the fundamental representation,
$h=$c$_2(R)/(k+N)$, c$_2(R)=(N^2-1)/2N$) carried out by the Wilson line.

The presence of ultraviolet divergences in the loop expansion of Chern-Simons
theory forces to regularize the theory and consequently to choose a
renormalization scheme. Certainly, from a perturbation theory point of view
all schemes are physically equivalent since they differ by a finite
renormalization which can be accomplished by adding  finite counterterms to
the action. However, one would like to know if there exist a scheme which
leads naturally to the exact result obtained by Witten. By naturalness we
understand a scheme in which the intermediate regularized action leads
after taking the limit in which the cutoff is removed to the exact result
obtained by Witten where the constant $k$ we started with (bare $k$) is the
same constant $k$ as the one appearing in the exact result. Certainly, this
concept of naturalness has only meaning in a theory like Chern-Simons theory
in which the beta function as well as the anomalous dimensions of the
elementary fields vanish at any order in perturbation theory
\REF\singer{S. Axelrod and I.M. Singer, MIT preprint, October
1991} [\bc,\dgs,\cpmdos,\singer]. This was the point of view taken in [\alr]
where  a scheme based on Pauli-Villars regularization seemed to be natural in
the sense discussed above. Indeed, the results obtained in [\alr] showed that a
choice of scheme of that type seems to lead to the shift  $k\to k+N$ which
appears in many of the equations corresponding to the exact result. The fact
that the origin of the shift is a quantum effect was first pointed out by
Witten
[\witCS] who showed its appearance using a gauge invariant regularization based
on the eta function. There are other schemes which lead to results in agreement
with [\alr,\witCS] as the one used in [\cpm]. All these schemes which seem to
provide at one loop an explanation of the origin of the shift share  the common
feature that the intermediate regularized action is gauge invariant.

So far all calculations involving quantum corrections using gauge
invariant regularized actions have concentrated on the effective action.
The fact that the quantum correction leads to an effective action whose
constant $k$ has been shifted indicates that one would observe such effect
when computing observables using those schemes. However, at present,
only indirect calculations of observables have been carried out taking into
account this quantum correction [\dror,\gmm]. In this paper we are going
to present the computation of the Wilson line corresponding to the unknot
in the fundamental representation of $SU(N)$ up to order $g^6$. This
calculation involves diagrams at two-loops for the two-point function
whose calculation in some scheme whose regularized action is gauge
invariant has to be carried out. We do not perform in this paper such a
two-loop calculation but we will show that agreement with the exact
result implies that there is no correction a two-loops for the two-point
function. The computation of the two-point function at two-loops using
the Pauli-Villars regularization plus higher derivatives proposed in [\alr]
is being carried out \REF\aalr{M. Alvarez, L. Alvarez-Gaum\'e, J.M.F.
Labastida and A.V. Ramallo, CERN preprint to appear} [\aalr]. So far we
have been able to prove that there is no need to introduce higher
derivative terms to regulate the theory at two loops and that a single
generation of Pauli-Villars is sufficient to render the two-loop graphs
finite. However, we have not finished the calculation of the finite part
that according to the results which we present in this work should be
zero to have agreement with the exact result, and, therefore, to be able to
consider the scheme based on Pauli-Villars as natural.

To end with this introduction we will reproduce here the result obtained
by Witten in [\witCS] for the framed unknot in the fundamental
representation of $SU(N)$ lying on $S^3$ with framing $n$.
The corresponding vacuum expectation value is,
$$
\langle W \rangle = {q^{N\over 2}-q^{-{N\over 2}}\over
q^{1\over 2}-q^{-{1\over 2}}} q^{n{N^2-1\over 2N}},
\eqn\uno
$$
where,
$$
q=\exp({2\pi i\over k+N}).
\eqn\dos
$$
Expanding \uno\ in terms of $1/k$ up to terms of order $1/k^3$ one finds,
$$
\eqalign{
\langle W \rangle =  N &+ {1\over k} i\pi n (N^2-1) \cr
&+{1\over k^2}\Big[-{\pi^2\over 6}N(N^2-1) -{\pi^2 n^2\over
2N}(N^2-1)^2-i\pi nN(N^2-1)\Big] \cr
&+ {1\over k^3}\Big[-{i\pi^3\over 6}n(N^2-1)^2-{i\pi^3\over
6N^2}n^3(N^2-1)^3 +\pi^2 n^2(N^2-1)^2 \cr
&\,\,\,\,\,\,\,\,\,\,\,\,\,\,\,\,\,
 +{\pi^2\over 3}N^2(N^2-1) +i\pi nN^2(N^2-1)\Big] \cr
&+ {\hbox{\rm O}}({1\over k^4}). \cr}
\eqn\cuatro
$$
Notice that in this expansion all terms containing a power of $\pi$
different that the power of $1/k$  are
originated by the fact that $k$ appears shifted into $k+N$ in \dos. In our
analysis we will show that those terms do indeed correspond to diagrams
which contain one-loop quantum corrections. Notice also that in the
standard framing ($n=0$) the series expansion has a simpler form. As a
consequence of our analysis we will be able to identify very simply all
diagrams which provide the framing dependence.
Actually, we will derive from a perturbation theory point of view the form
of the framing dependence of the vacuum expectation value of an arbitrary
knot. If, on the other hand, it turns to be correct the picture in which there
are only one-loop corrections (which just account for the shift $k\to k+N$)
one could extract all the effects due to framing and therefore one would
be  left with a series of diagrams which constitute the building blocks of
the knot invariant. These building blocks lead to topological invariants
which after considering them as the coefficients of a power series build
the knot invariants leading to the Jones polynomials \REF\jones{V.F.R.
Jones, {\sl Bull. Am. Math. Soc.} {\bf 12} (1985) 103,
{\sl Ann. Math.} {\bf 126} (1987) 335}  [\jones]
\REF\homfly{P. Freyd, D. Yetter, J. Hoste, W.B.R. Lickorish, K. Millet and
A. Ocneanu, {\sl Bull. Am. Math. Soc.} {\bf 12} (1985) 239}
\REF\kauf{L.H. Kauffman, {\sl Trans. Amer. Math. Soc.} {\bf 318}
(1990) 417} and its cousins [\homfly,\kauf]. We will discuss in more detail
this picture of the perturbation theory series expansion in our concluding
remarks.

In this paper we will consider the three dimensional manifold as $R^3$
which allows us to identify the corresponding observables to the ones in
$S^3$. We will not discuss the effect of the framing of the
three-dimensional manifold from a perturbation theory point of view. A
good discussion of this point can be found in [\dror].

The paper is organized as follows. In sect. 2 we define the regularized
Chern-Simons gauge theory using Pauli-Villars fields which we claim to
correspond to a natural scheme in the sense discussed above. In sect. 3
we compute \cuatro\ in perturbation theory the vacuum expectation value of the
unknot carrying the fundamental representation of $SU(N)$ up to order $g^6$.
In
sect. 4 we will identify all the framing dependence of the vacuum expectation
value of a knot and we will show its factorization in the form predicted by
Witten. Finally, in sect. 5 we state our conclusions and make some final
remarks
   .
Several appendices deal with our conventions and with the proof of some results
which are used in sects. 3 and 4.

\endpage
\chapter{Perturbative Chern-Simons gauge theory}

In this section we will define Chern-Simons gauge theory from a
perturbation theory point of view. This is carried out in two steps.
First a gauge fixing is performed. Second, after analyzing the ultraviolet
behavior of the theory a regularized action using Pauli-Villars fields is
provided. Let us consider an $SU(N)$ gauge connection $A_\mu$ on a
boundaryless three-dimensional manifold $M$ and the following Chern-Simons
action, $$
S(A_\mu) = {k\over 4\pi}\int_M \tr(A\wedge d A + {2\over 3} A\wedge
A\wedge A),
\eqn\cinco
$$
where $k$ is an arbitrary positive integer\foot{A negative $k$ will
change the $\epsilon$ prescription in the perturbative series expansion
leading to a shift of $k$ at one loop with the opposite sign. With a
negative $k$ one makes connection with the exact result (1.1) after
replacing $q\to q^*$.} and ``Tr" denotes the trace in the fundamental
representation of $SU(N)$ (normalized in such a way that
$\tr(T^a T^b) = -{1\over 2} \delta^{ab}$). A summary of our
group-theoretical conventions is contained in Appendix A.
In defining the theory from a perturbation theory point of view we must
give a meaning to vacuum expectation values of operators, \ie, to
quantities of the form,
$$
\langle {\cal O} \rangle = {1\over Z}
\int [D A_\mu] {\cal O} \exp\Big(iS(A_\mu)\Big),
\eqn\seis
$$
where $Z$ is the partition function,
$$
Z=\int [D A_\mu]  \exp\Big(iS(A_\mu)\Big).
\eqn\seisp
$$
The operators entering \seis\ are gauge invariant operators which do not
depend on the three-dimensional metric. These operators are knots, links
and graphs \REF\witGR{E. Witten \journal\np&B322(89)629} [\witCS,\witGR].
The first issue in defining \seis\ is to take care of the gauge fixing.
Indeed, the exponential in \seis\ is invariant under gauge
transformations of the form,
$$
A_\mu \to h^{-1}A_\mu h + h^{-1}\partial_\mu h,
\eqn\siete
$$
where $h$ is an arbitrary continuous map $h:M\to SU(N)$. Before carrying
out the gauge fixing let us redefine the constant $k$ and the field
$A_\mu$ in such a way that the action \cinco\ becomes standard from a
perturbation theory point of view. We define,
$$
g = \sqrt{{4\pi\over k}}.
\eqn\ocho
$$
Then, after rescaling the gauge connection,
$$
A_\mu \to g A_\mu,
\eqn\nueve
$$
one obtains the following Chern-Simons action,
$$
S'(A_\mu) = \int_M \tr(A\wedge d A + {2\over 3} g A\wedge
A\wedge A).
\eqn\diez
$$
This form of the action contains the standard 1/2 factor for the kinetic
term after using (A3). From now on we will restrict ourselves to the case
in which the three-dimensional manifold $M$ is $R^3$ which is the
simplest case to treat from a perturbation theory point of view. Though
\diez\ is metric independent, we will be forced to introduce a metric in
carrying out the gauge fixing. We will assume that this metric has
signature $(1,-1,-1)$.

Our gauge choice will be the same as the one taken in [\alr]. The
Lorentz-like gauge condition $\partial^\mu A_\mu=0$ is imposed using the
standard Fadeev-Popov construction which leads to the following
action to be added to \diez,
$$
S_{{\hbox{\sevenrm gf}}}(A_\mu,c,\bar c,\phi)=
\int\tr\Big(2\bar c\partial_\mu D^\mu c- 2\phi\partial_\mu A^\mu
-\lambda \phi^2\Big),
\eqn\once
$$
where $\phi$ is the Lagrange multiplier which imposes the gauge
condition, $c$ and $\bar c$ are the Fadeev-Popov ghost, and $\lambda$ is
a gauge fixing parameter. In \once, $D_\mu$ is the covariant derivative,
$D_\mu c = \partial_\mu c + g[A_\mu,c]$. The action \diez\ as well as the
gauge-fixing  action \once\ are invariant under the following BRST
transformations,
$$
sA_\mu = D_\mu c,\,\,\,\,\,\, sc=-cc,\,\,\,\,\,\,
 s\bar c =\phi,\,\,\,\,\,\, s\phi=0.
\eqn\doce
$$
The field $\phi$ can be integrated out easily providing the following
functional integral for vacuum expectation values as the ones in \seis:
$$
\langle {\cal O} \rangle = {1\over Z}
\int [D A_\mu Dc D\bar c] {\cal O} \exp\Big(iI(A_\mu,c,\bar c)\Big),
\eqn\trece
$$
where,
$$
I(A_\mu,c,\bar c)=\int\tr\Big(\epsilon^{\mu\nu\rho}(A_\mu\partial_\nu
A_\rho+{2\over 3}A_\mu A_\nu A_\rho)- \lambda^{-1}A_\mu
\partial^\mu\partial^\nu A_\nu + 2 \bar c \partial_\mu D^\mu c\Big).
\eqn\catorce
$$
Of course, $Z$ in \trece\ is  appropriately defined taking into account the
gauge fixing. The quantities obtained in \trece\ are independent of the
value of $\lambda$. In order to avoid the presence of infrared
divergences we will work in  the Landau gauge in which $\lambda=0$

The perturbative series expansion which one obtains from \trece\ and
\catorce\ possesses some divergences which need to be regularized.
The analysis of the nature of these divergences was carried out in [\alr]
by performing the corresponding power counting. There are many ways to
regularize these divergences giving physically equivalent results. In
this work we will follow the regularization procedure introduced in
[\alr], \ie, we will use a gauge invariant regularization based on
the introduction of Pauli-Villars fields and, if needed,
higher-derivative terms. This seems to provide a scheme which is natural
in the sense explained in sect. 1. Further work have shown [\aalr]
that there is no need to introduce higher-derivative terms. The
Pauli-Villars fields which one introduces to regulate at one loop seem to
be sufficient to render the theory finite to any order. Of course, after
the gauge fixing has been performed, when talking about a gauge invariant
regularization we mean a regularization which preserves the
BRST symmetry \doce.

Following [\alr] we introduce Pauli-Villars fields $A_\mu^{(j)}$,
$c^{(i)}$ and $\bar c^{(i)}$, $j=1,...,J$ and $i=1,...,I$. The
regularized functional integral takes the form,
$$
\langle {\cal O} \rangle = {1\over Z}
\int [D A_\mu Dc D\bar c] {\cal O} \exp\Big(iI(A_\mu,c,\bar c)\Big)
\Big(\prod_{j=1}^J {\det}^{-b_j/2}{\cal A}_j\Big)
\Big(\prod_{i=1}^I {\det}^{c_i}{\cal C}_i\Big),
\eqn\quince
$$
where, of course, the same type of regularization is used for $Z$, and
all the dependence on the Pauli-Villars fields is contained in
the determinants,
$$
\eqalign{
{\det}^{-1/2}{\cal A}_j &=
\int [DA_\mu^{(j)}] \exp\Big(
i\int\tr(\epsilon^{\mu\nu\rho}A_\mu^{(j)}D_\nu A_\rho^{(j)}+M_j
A_\mu^{(j)}A^{(j)\mu})\Big) \cr
{\det}{\cal C}_i &= \int[D\bar c^{(i)}Dc^{(i)}]
\exp\Big(2i\int\tr(\bar c^{(i)}D_\mu D^{\mu}c^{(i)}-m_i^2
\bar c^{(i)} c^{(i)})\Big).\cr}
\eqn\dseis
$$
The masses entering into the determinants in \dseis\ as well as the
integers $b_j$, $j=1,...,J$ and $c_i$, $i=1,...,I$ are the regulating
parameters. The relative values of these masses and these integers are fixed
to make the theory finite in the limit in which the common scale of the
masses $\Lambda$ becomes large. In [\alr] was shown that the following
choice makes the theory finite at one loop,
$$
\eqalign{
&\sum_{j=1}^J b_j =1, \,\,\,\,\,\,\,
\sum_{j=1}^J {b_j\over M_j} = 0, \,\,\,\,\,\,\,
\sum_{j=1}^J {b_j\over M_j^2} =0,\cr
&I=J,  \,\,\,\,\,\,\,
c_j={1\over 2}b_j,  \,\,\,\,\,\,\,
m_j= M_j. \cr}
\eqn\dseisp
$$
We conjecture that the limit
$\Lambda\to\infty$ of \quince\ with the choice \dseisp\ generates the same
values for the observables of the theory (once the ambiguities originated
at coincidence points of products of operators are handled as shown in
the next section) as the ones in the exact result obtained by Witten
[\witCS]. The results presented in this paper and in [\aalr] provide
certain evidence  towards the validity of this conjecture.

As shown in [\alr] the regularized action entering \quince\ is BRST
invariant. Indeed, defining the BRST transformations of the Pauli-Villars
fields as just gauge transformations of fields transforming in the adjoint
representation whose gauge parameter is the ghost field $c$,
$$
sA_\mu^{(j)} = [ A_\mu^{(j)},c], \,\,\,\,\,\,\,
s\bar c^{(i)} = \{\bar c^{(i)},c\}, \,\,\,\,\,\,\,
s c^{(i)} = \{c^{(i)},c\}, \,\,\,\,\,\,\,
\eqn\dsiete
$$
it is simple to prove that the determinants entering \quince\ are BRST
invariant.

To end this section let us summarize the Feynman rules of the theory as
well as the one-loop results obtained in [\alr]. They will
become very useful in the next section where the Wilson line corresponding
to the unknot will be computed to order $g^6$. We will work in space-time
space. The two basic Feynman rules entering our calculations are summarized
in \FIG\launo{Basic Feynman rules of the theory.} Fig. \launo. In
particular, the propagator associated to the gauge field takes the form,
$$
\Sigma^{\mu\nu}_{ab}(x,y)={i\over 4\pi}\delta_{ab}
\epsilon^{\mu\rho\nu}{(x-y)_\rho\over |x-y|^3}.
\eqn\diocho
$$
We do not give the Feynman rules corresponding to ghost and Pauli-Villars
fields since these fields only enter in loops and we will take the results
obtained in [\alr] for one-loop Green functions. These results are
summarized in \FIG\lados{Two-point function and three-point function at
one loop.} Fig. \lados.

\endpage
\chapter{Unknot to order $g^6$.}

In this section we will compute the vacuum expectation value of the
Wilson line corresponding to the unknot in the fundamental representation
of $SU(N)$ using the functional integral defined in \quince. This
calculation will provide the tools and methods to analyze general
features of the perturbative series expansion of vacuum expectation
values of knots as the one considered in the next section. Taking into
account the rescaling \nueve, the operator ${\cal O}$ entering in \quince\
has the form, $$ W = \tr \Big({\hbox{\rm P}}\ex^{g\oint A}\Big),
\eqn\dinueve $$
where the trace is taken over the fundamental representation and P
denotes  path-ordered product. This choice of sign in the exponential leads to
the convention (A1). The contour integral in \dinueve\ corresponds to any path
diffeomorphic to the unknot. To compute the vacuum expectation value of this
operator in perturbation theory we have to consider all  diagrams which are not
vacuum diagrams since, as shown in \quince, we consider normalized vacuum
expectation values, \ie, in \quince, the functional integration where the
operator is inserted is divided by the partition function $Z$.
The expansion of the path-ordered exponential in \dinueve\ reduces the
calculation to certain integrals of $n$-point functions. These $n$-point
functions need to be computed perturbatively up to certain order. We will
use the standard Feynman diagrams to denote these $n$-point functions.
To denote the contour integral we will attach their $n$-points by a
circle. For convenience, let us express the perturbative series
corresponding to the vacuum expectation value of \dinueve\ as,
$$
\langle W \rangle = \sum_{i=0}^\infty w_{2i} g^{2i}.
\eqn\veinte
$$
Clearly, to order $g^0$ the computation of the vacuum expectation
value of \dinueve\ reduces to the trace of the unit operator in the
fundamental representation which is just $N$,
$$
w_0 = N,
\eqn\vuno
$$
in agreement with \cuatro.
Higher orders up to $g^6$ will be computed in the following subsections.

\section{Order $g^2$}

To this order, since there is a factor $g$ in the exponential  \dinueve\
there is only one diagram which just involves the propagator \diocho.
This diagram is shown in \FIG\latres{Diagrams corresponding to $g^2$. Thick
lines represent the Wilson line while thin lines refer to Feynman diagrams.}
Fig
   .
\latres. Its contribution to the perturbative series \veinte\ is just,
$$
\eqalign{
w_2 = &  \tr(T^aT^b)\oint dx^\mu\int^x dy^\nu {i\over 4\pi}
\delta_{ab}\epsilon^{\mu\rho\nu} {(x-y)_\rho\over |x-y|^3}\cr
=&{1\over 2}\tr(T^aT^b)\oint dx^\mu\oint dy^\nu {i\over 4\pi}
\delta_{ab}\epsilon^{\mu\rho\nu} {(x-y)_\rho\over |x-y|^3}.\cr }
\eqn\vdos
$$
To perform the step carried out in obtaining the second expression for
$w_2$ one must first realize that the integration is well defined and
finite, and symmetric under the interchange $x^\mu \leftrightarrow y^\nu$.
Notice that although it seems that there are singularities at coincident
points, a careful analysis of the integral shows that this is not the case
[\pol,\gmm]. However, from a quantum field
theory point of view the quantity entering \vdos\ is not well defined. The
reason is that among the points of integration there are points where one
is using quantities like $\langle A_\mu(x)A_\nu(x) \rangle$ which are not
well defined from a field theory point of view. One could add a finite part
at those coincident points making the integration ambiguous. As shown in
[\gmm] there is way to solve this ambiguity providing a procedure which is
metric independent as it would be desirable from the point of view of
topological field theory. The idea is to introduce an unit vector $n^\mu$
normal to the path of integration and consider the path corresponding to
$y^\nu$ as the one constructed by $y^\nu = x^\nu + \varepsilon n^\nu$. The
resulting integral depends on the choice of $n^\mu$ and it corresponds to
the the Gauss integral which can be normalized such that its value is an
integer $n$,
$$ n={1\over 4\pi} \oint dx^\mu {\oint}' dy^\nu \epsilon^{\mu\nu\rho}
{(x-y)_\rho\over |x-y|^3}.
\eqn\vtres
$$
In this equation the prime denotes that the second path is slightly separated
from the first path as dictated by the unit vector $n^\mu$. Often we will refer
to this situation as saying that $x$ runs over the knot and $y$ over its frame.
The integer value
$n$ is the linking number of the two non-intersecting paths. In general,
the perturbative expansion of the Wilson line will possess terms
containing the ambiguity discussed here. From a field theory point of
view, one may detect the presence of this ambiguity just observing if
in the integrations of products of operators one is integrating over
coincident points. Fortunately, this seems to happen only when the two end
points of a propagator may get together (``collapsible" propagator). It turns
out that the three-point function  possesses milder
singularities than the propagator at coincident points and it does not
introduce any ambiguity. We will discuss in more detail this feature
in the next section.

Using \vtres\ and (A3) one finds for $w_2$ in \vdos,
$$
w_2= {i\over 4}n(N^2-1),
\eqn\vcuatro
$$
which is in agreement with  \cuatro\ after taking into account that
$g^2 =4\pi/k$.

\section{Order $g^4$}

The diagrams contributing to this order are depicted in
\FIG\lacuatro{Diagrams corresponding to $g^4$.}
Fig. \lacuatro. It is at this order where the first appearance of a
diagram involving quantum corrections is present. Namely, diagram
$a$ of Fig. \lacuatro\ contains the full two-point function at one
loop. This two-point function was computed in [\alr] in the scheme
adopted in this paper. The result obtained there has been summarized
in Fig. \lados. Taking into account that result and the
previous calculation leading to $w_2$ we can write very simply the
contribution of diagram $a$ of Fig. \lacuatro\ to this order,
$$
w^{(a)}_4 = -{N\over 4\pi}w_2 = -{i \over 16\pi} n N (N^2-1).
\eqn\vcinco
$$
Notice that this contribution corresponds to the last one at order
$1/k^2$ in \cuatro. This term in \cuatro\ is such that the power of $\pi$
and the power of $1/k$ are different and therefore corresponds to the type
of terms which are in the expansion of $\langle W\rangle$ because $k$
appears shifted into $k+N$ in the exact result \uno. This is the first case in
which we will observe that a diagram present because of the existence of
quantum corrections gives a contribution which corresponds to the one
originated by the shift present in the exact result.

The contribution of diagrams $b$, and $c_1$, $c_2$ and $c_3$ of Fig.
\lacuatro\ has been analyzed in detail in [\gmm,\dror]. We will use here
their results and we will make a series of remarks which will be useful in
computations at higher order. The contribution from $b$ is,
$$
\eqalign{
w_4^{(b)} = & \tr(T^aT^bT^c)\oint dx^\mu \int^x dy^\nu \int^y dz^\rho
\int d^3\omega \Big( (-i)f^{abc}\epsilon_{\nu_1\nu_2\nu_3} \cr &
\,\,\,\,\,\,\,
{i\over 4\pi}\epsilon^{\mu\rho_1\nu_1}{(x-w)_{\rho_1}\over
|x-w|^3} {i\over 4\pi}\epsilon^{\nu\rho_2\nu_2}{(y-w)_{\rho_2}\over |y-w|^3}
{i\over 4\pi}\epsilon^{\rho\rho_3\nu_3}{(z-w)_{\rho_3}\over |z-w|^3}\Big)
\cr = &
{1\over 8} N (N^2-1) \rho_1(C),\cr}
\eqn\vseis
$$
where we have used (A1) and (A3) and,
$$
\eqalign{
\rho_1(C)={1\over 32\pi^3}
\oint dx^\mu \int^x dy^\nu \int^y dz^\rho
\int d^3\omega \Big( &\epsilon^{\mu\nu_1\rho_1}\epsilon^{\nu\nu_2\rho_2}
\epsilon^{\rho\nu_3\rho_3}\epsilon_{\nu_1\nu_2\nu_3} \cr &
{(x-w)_{\rho_1}(y-w)_{\rho_2}(z-w)_{\rho_3} \over
|x-w|^3 |y-w|^3 |z-w|^3}\Big). \cr}
\eqn\vsiete
$$
This quantity has a special significance which we will discuss
after analyzing the contribution from the rest of diagrams at this order.
The argument of $\rho_1(C)$, $C$, is the  integration path. Notice that the
integration entering $\rho_1(C)$ does not possess any ambiguity due to
the presence of products of operators at coincident points and therefore
it is framing independent.  The reason why ambiguities are not present is
that coincident points occur pairwise, \ie, the three endpoints never
get together in the integration, and singularities associated to this
case are too mild to introduce ambiguities.
Of course, this assertion needs a careful proof which indeed has been
carry out indirectly in [\gmm,\dror]. Form a quantum field theory point of
view, it seems plausible and we will think about it as a general feature
of the perturbative series expansion.  For the unknot the
quantity $\rho_1(C)$
was computed in [\gmm] obtaining the result,
$$
\rho_1|_{\hbox{\sevenrm
unknot}} = -{1\over 12}. \eqn\vocho
$$
Taking into account this value, the contribution from
diagram $b$ of Fig. \lacuatro\ has the form,
$$ w_4^{(b)} = -{1\over 96} N
(N^2-1), \eqn\vnueve
$$
which is just the first term of order $1/k^2$ in \cuatro\ after taking
into consideration that $g^2=4\pi/k$.

We are left with the contributions from diagrams $c_1$, $c_2$ and $c_3$
of Fig. \lacuatro.
Diagrams $c_1$ and $c_2$ give the same contribution. However, diagram
$c_3$ has an entirely different nature. On the one hand, notice that
diagram $c_3$ does not possess ambiguities. The endpoints of a
propagator never get together since they always enclose an endpoint of
another propagator. This means in particular that the contribution from
such a diagram is framing independent. In addition, the group factor
from this diagram is different than the one from the other two diagrams.
Non-planar diagrams as $c_3$ possess different group factors than the
corresponding planar ones. In general, using (A1) the group factor of a
non-planar diagram can be decomposed in a part containing the same
structure as the planar one plus another contribution. Namely using (A1)
one finds,
$$
\eqalign{
\tr(T^aT^bT^aT^b) = & \tr(T^aT^aT^bT^b) + f^{abc}\tr(T^aT^cT^b) \cr
= & {(N^2-1)^2\over 4N} - {1\over 4}N(N^2-1). \cr}
\eqn\treinta
$$
the first group factor has the same form as the group factors of diagrams
$c_1$ and $c_2$ and we will consider all three contributions together.
Actually it is simple to realize that the resulting expression once the
three contributions are taken into account possesses an integrand that is
symmetric. This allows to enlarge the integration region symmetrically
and divide by a factor 4!. On the other hand the contribution due to the
second group factor in \treinta\ is proportional to,
$$
\rho_2 (C) = {1\over 8\pi^2}\oint dx^\mu\int^x dy^\nu \int^y dz^\rho
\int^z d\omega^\tau \epsilon^{\mu\sigma_1\rho}\epsilon^{\nu\sigma_2\tau}
{(x-z)_{\sigma_1}\over |x-z|^3}{(y-w)_{\sigma_2}\over |y-w|^3},
\eqn\tuno
$$
which vanishes for the case in which the contour $C$ can be contained in
a plane as it is the case for the unknot. Therefore, the contribution
from diagrams $c_1$, $c_2$ and $c_3$ of Fig. \lacuatro\ takes the form,
$$
\eqalign{
w_4^{(c)}= &-{(N^2-1)^2\over 4N}{3\over 4!}
\oint dx^\mu\oint dy^\nu\oint dz^\rho\oint dw^\tau \Big(
{\epsilon^{\mu\sigma_1\nu}\over 4\pi}{(x-y)_{\sigma_1}\over |x-y|^3}
{\epsilon^{\rho\sigma_1\tau}\over 4\pi}{(z-w)_{\sigma_2}\over |z-w|^3}
\Big)\cr =&-{n^2(N^2-1)^2\over 32N},\cr}
\eqn\tdos
$$
where in the last step we have used \vtres. This contribution is just the
remaining one at order $1/k^2$ in the expansion \cuatro. Therefore, to this
order we have full agreement between the exact result and the
perturbative calculation. Notice  that to achieve this we have defined
products of operators at coincident points in  a very precise manner. We
have argued that the ambiguity in those products only produces a relevant
effect when the points of coincidence are joined by a propagator.
Coincidence of end-points which belong to a connected part of an
$n$-point function with $n>2$ does not introduce any ambiguity. One may
verify that the singularities appearing when $n>2$ are milder than in the
case $n=2$ to justify in certain sense that assertion. However, a
complete proof of it would be desirable. For the case of $\rho_1(C)$ and
$\rho_2(C)$, it has been shown [\gmm,\dror] that both are framing
independent, in agreement with our statement. Their sum must therefore
correspond to a knot invariant. In fact, it was shown in [\gmm] that
$$
\rho(C) = \rho_1(C) + \rho_2(C)
\eqn\ttres
$$
can be identified with the second coefficient of the Alexander-Conway
polynomial. In general, the picture that emerges from the perturbative
calculation is that the connected $n$-point functions, $n>2$, constitute the
main building blocks of the knot invariant \uno. This building blocks are knot
invariants and build the perturbative series leading to \uno. The
two-point function takes care of the framing (planar contribution) and of
some corrections to the connected $n$-point functions, $n>2$, as
$\rho_2(C)$ above (non-planar contribution). We will see how these facts
are realized at  next order in perturbation theory. Their general features
will be discussed in sect. 4.

\section{Order $g^6$}

This is the first order where a two-loop diagram takes place.
The diagrams contributing to this order are represented in \FIG\lacinco{Part of
the diagrams corresponding to $g^6$.} Fig. \lacinco\ and \FIG\laseis{Rest of
diagrams corresponding to $g^6$.}Fig. \laseis. Diagram $a_1$ involves the full
two-loop one particle irreducible two-point function. This quantity has not
been
computed yet in the regularization scheme considered in this paper. One of the
aims of this work is to demonstrate that it must vanish in order to have
agreement with the exact result \uno. We will compute in this section all other
contributions at this order and we will prove that they generate all the terms
at order $1/k^3$ in \cuatro.

The contribution from diagram $a_2$ is straightforward after using
the expression in Fig. \lados. It turns out,
$$
w_6^{(a_2)}= {i\over 64\pi^2}nN^2(N^2-1).
\eqn\tcuatro
$$
This contribution corresponds to the last term at order $1/k^3$ in
\cuatro. Notice that this is one of the terms where the power of $\pi$ is
different than the power of $1/k$ and therefore is shift related. The
other diagrams containing one-loop corrections are $b$, $c_1$, $c_2$ and
$c_3$, and $d_1,...,d_6$ of Fig. \lacinco. The contribution from these diagrams
are simple to compute using the form of the one-particle irreducible diagrams
in
Fig. \lados, and the results of the previous order. From diagram $b$ one
finds,
$$
w_6^{(b)} = {1\over 32\pi}N^2(N^2-1)\rho_1
\eqn\tcinco
$$
while, similarly, from diagrams $c_1$, $c_2$ and $c_3$, which all give
the same contribution,
$$
w_6^{(c)} = -{3\over 32\pi}N^2(N^2-1)\rho_1.
\eqn\tseis
$$
Finally, after rearranging the group factors as in \treinta, the
contribution from diagrams $d_1,...,d_6$ is,
$$
w_6^{(d)} = {1\over 64\pi}n^2(N^2-1)^2 -{1\over 16\pi}N^2(N^2-1)\rho_2.
\eqn\tsiete
$$
Collecting all the contributions and using \vocho\ and the fact that for
the unknot $\rho_2=0$ one finds,
$$
\eqalign{
w_6^{(b)} + w_6^{(c)} +w_6^{(d)} =&
-{1\over 16\pi}N^2(N^2-1)(\rho_1+\rho_2) + {1\over 64\pi}n^2(N^2-1)^2\cr
=& {1\over 192\pi}N^2(N^2-1) + {1\over 64\pi}n^2(N^2-1)^2, \cr}
\eqn\tocho
$$
which correspond to the other two contributions in \cuatro\ (all except the
last one) whose power of $\pi$ does not coincide with the power of $1/k$.

The rest of the diagrams contributing at this order do not contain loop
corrections and are depicted in Fig. \laseis. Diagrams $e_1$, $e_2$ and $e_3$
of
Fig. \laseis\ involve the tree-level four-point function. Clearly, the first
two
diagrams are planar  and identical while the third one is non-planar. This
third diagram, $e_3$, possesses the group factor,
$$
\tr(T^aT^bT^cT^d)f^{ace}f^{ebd},
\eqn\tnueve
$$
which, as shown in Appendix A, vanishes (see equation (A7)).
The other two diagrams, $e_1$ and  $e_2$ of Fig. \laseis, which are the same,
vanish for the case of the unknot as it is shown in Appendix B.

Let us compute the contribution form the ten diagrams $f_1,...,f_{10}$ of
Fig. \laseis. These diagrams can be divided in planar and non-planar
ones. As in previous cases, non-planar diagrams possess group factors
which decompose into the group factors of the planar ones plus an
additional contribution. Indeed, from a diagram like $f_6$ the group
factor is,
$$
\tr(T^aT^aT^bT^cT^d)f^{bcd} = -{1\over 8} (N^2-1)^2,
\eqn\cuarenta
$$
while from a diagram like $f_2$ the group factor is,
$$
\tr(T^aT^bT^cT^dT^b)f^{acd} = -{1\over 8} (N^2-1)^2 + {1\over
8}N^2(N^2-1).
\eqn\cuno
$$

Non-planar diagrams of this type do not contribute for the case in which the
Wilson line corresponds to the unknot. This can be shown writing explicitly the
integration involved or using the lemma below. The main idea behind the
argument based on that lemma is that non-planar diagrams of the type under
consideration are framing independent so one can choose any framing to compute
it. For the unknot it is simple to realize that choosing a framing which is
contained in the same plane as the unknot the integrand vanishes trivially.
Before stating and proving this lemma let us define ``free'' propagators
as the ones that have both endpoints on the knot.

{\bf Lemma.} {\sl
Every framing independent diagram of the unknot containing a free
propagator is zero.}

{\sl Proof.} Let us place the unknot $C$ in
a plane. Being the diagram framing independent, choose a frame $C_f$
coplanar to it. The diagram contains the part corresponding to the free
propagator, $$
 \oint  \cdots dx^\mu  dy^\nu \epsilon_{\mu \rho \nu}
{(x-y)^{\rho} \over |x-y|^3} \cdots
\eqn\cuno
$$
where $dx^\mu \in C$ and $dy^\nu \in C_f$. Due to the coplanarity,
the previous term is a $3 \times 3$ determinant whose rows are
linearly dependent. Then, it is zero and the lemma is proved.
This result is very powerful once all the framing independent diagrams of
the perturbative series expansion are identified. The theorem stated in the
next section allows to characterize very simply all those diagrams.
As we will discuss there, it turns out that those diagrams are the ones not
containing collapsible propagators. Thus, using the lemma above, we  conclude
that the only non-vanishing diagrams contributing to the perturbative series
expansion of the unknot are the ones with no free propagators.

We are left with
planar diagrams of type $f$ in Fig. \lacinco. Actually, it will be much
more convenient to consider the whole set of the ten diagrams all with the
same group factor \cuarenta. The reason for this is that then one can show
the factorization of the contribution into a product of contributions of
the type appearing in Fig. \latres\ times contributions of the type $b$ in
Fig. \lacuatro. This phenomena of factorization is  general
for diagrams with disconnected one-particle irreducible subdiagrams.
Indeed, in Appendix C we show the general form of the factorization
theorem.  The result of applying this theorem for diagrams $f_1,...,f_{10}$
of Fig. \laseis\
is explained as an example in Appendix C. It turns out that it can be written
as the following product:
 $$
\eqalign{& w_6^{(f)} = {-i \over 8} (N^2-1)^2 g^6 {1 \over 4\pi}{1 \over
2}\oint dx_1^{\alpha_1}\oint dx_2^{\alpha_2}\epsilon^{\alpha_1
\alpha_2 \alpha}{(x_1-x_2)_{\alpha} \over |x_1-x_2|^3}{1 \over
64\pi^3}\times\cr
& \oint\! dx_3^{\alpha_3}\int^{x_3}\! dx_4^{\alpha_4}
\int^{x_4}\! dx_5^{\alpha_5}\int\! d^3z \epsilon^{\alpha_3 \alpha_6 \gamma}
\epsilon^{\alpha_4 \alpha_7 \delta}\epsilon^{\alpha_5 \alpha_8 \beta}
\epsilon_{\alpha_6 \alpha_7 \alpha_8}{(z-x_3)_{\gamma} (z-x_4)_{\delta}
(z-x_5)_{\beta} \over |z-x_3|^3|z-x_4|^3|z-x_5|^3}, \cr}
\eqn\cunop
$$
\ie, a product of a linking number times  $\rho_1$,
$$
w_6^{(f)} = {i\over 32}n(N^2-1)^2\rho_1 =
-{i\over 384}n(N^2-1)^2.
\eqn\cdos
$$
In obtaining \cdos\ we have used  \vcuatro, \vseis\ and \vocho. This
contribution is just the first one at order $1/k^3$ in \cuatro\ after
using the fact that $g^2=4\pi/k$.
This procedure of using the lemma plus the factorization theorem of Appendix
C is a general feature of the unknot. In general, for an arbitrary knot, the
factorization theorem would force us to overcount diagrams giving additional
contributions. However, for the unknot all those contributions vanish.

%

To complete the  perturbative computation at order $g^6$ we are left with
diagrams $g_1,...,g_{15}$ of Fig. \laseis. Again these diagrams can be divided
i
   n
planar and non-planar ones. However, now the non-planar ones can be divided in
three groups depending on the number crossings. The
group factor decomposes differently in each group.
 A given diagram produces an additional group factor
for each uncrossing needed to make it planar. If the
group factor of the planar diagrams, $g_1$ to $g_5$ is
$$ \tr(T^aT^aT^bT^bT^cT^c)
= -{1\over 8N^2} (N^2-1)^3, \eqn\ctres
$$
the group factor of diagrams $g_6,...,g_{11}$, which are of the first
type, takes the form,
$$
\tr(T^aT^aT^bT^cT^bT^c) = -{1\over 8N^2} (N^2-1)^3 +{1\over 8}(N^2-1)^2,
\eqn\ccuatro
$$
where we have used simply (A1). The diagrams of the second type are
$g_{12}$, $g_{13}$ and $g_{14}$, which similarly generate the following
group factor,
$$
\tr(T^aT^bT^aT^cT^bT^c) = -{1\over 8N^2} (N^2-1)^3 +{1\over
8}(N^2-1)^2 +{1\over 8}(N^2-1)
\eqn\ccinco
$$
Finally, diagram $g_{15}$ generates,
$$
\tr(T^aT^bT^cT^aT^bT^c) = -{1\over 8N^2} (N^2-1)^3 +{1\over
8}(N^2-1)^2 +{1\over 4}(N^2-1)
\eqn\cseis
$$

Of all three types of group factors only the first one contributes in the
case of the unknot. To the second group factor, ${1\over
8}(N^2-1)^2$, there are contributions from the last 10 diagrams. Using the
factorization theorem of Appendix C one finds that this
contribution is proportional to $\rho_2$ (diagram $c_3$ in Fig. \lacuatro)
and therefore vanishes.  We have rearranged the group factors in order to get
th
   e
right weights which make explicit the factorization of $\rho_2$. To the third
gr
   oup
factor, ${1\over 8}(N^2-1)$, there are contributions from the last 4 diagrams
wh
   ich
can be shown explicitly to vanish for the case of the unknot. We are left with
t
   he first
group factor, $-{1\over 8N^2} (N^2-1)^3$. There are contributions from all
diagrams. One can use the factorization theorem of Appendix C to write this
contribution as a product of contributions of the type shown in Fig. \latres.
Us
   ing
\vcuatro\ one then finds, $$
w_6^{(g)} = - {i\over 384 N^2} n^3 (N^2-1)^3,
\eqn\csiete
$$
which indeed corresponds to the second contribution at order $1/k^3$ in
\cuatro. The calculation is described in some detail as an example in Appendix
C. This was the only contribution left to be obtained from the perturbative
series expansion. The agreement found between the two results shows that the
contribution from diagram $a_1$ of Fig. \lacinco\ must be zero. This implies
that the one-particle irreducible diagram corresponding to the two-point
function must vanish at two loops.

\endpage

\chapter{Factorization of the framing dependence}

In this section we state a theorem about the framing independence of
diagrams which do not contain one-particle irreducible subdiagrams
corresponding to two-point functions whose endpoints could get together.
This theorem refers to any kind of knot. Before stating it,
 some remarks are in order. Let us consider an arbitrary
diagram whose $n$ legs are attached to $n$ points on the knot.
The resulting integral
runs over these points on the knot in a given order, $i_1<i_2<\cdots <i_n$.
Suppose that our diagram has a propagator with endpoints attaching two
consecutive points, say $i_1$ and $i_2$. Remember that the path ordered
integration will make $i_1 \to i_2$, and that the propagator is singular in
that case. Albeit this singularity exists, the integral is finite but
shape-dependent, as is well-known. The results for a circumference and for
an ellipse are different and then it is not a topological invariant.
The way out of this difficulty is the introduction of
framings [\pol,\witCS,\gmm]. When the propagator connects the knot and the
frame, the resulting integral is the linking number of the frame around the
knot, and this is a topological invariant. This suggests that the framing is
relevant only when there are collapsible (free propagators
whose  endpoints may get together upon integration) propagators. This is
the idea behind this theorem. Its statement is:
\vskip .5cm
{\bf Theorem.}  {\sl A diagram gives a framing dependent
contribution to the perturbative expansion of the knot
if and only if it contains at least one collapsible propagator.
Moreover, the order of $n$ in its contribution, the linking number, equals the
number of collapsible propagators}.
\vskip .5cm
Diagrams $b$ and  $c_3$ of Fig. \lacuatro, $e_1, e_2$ and $e_3$,  $f_1$ to
$f_5$,  and $g_{12}$ to $g_{15}$ of Fig. \laseis\  are examples of framing
independent diagrams. Diagrams $a$, $c_1$ and $c_2$ of Fig. \lacuatro, and
 $f_6$ to $f_{10}$, and $g_1$ to $g_{11}$  of Fig. \laseis\ are
examples of framing dependent ones.

Although we have no rigorous proof of this theorem, we do have
results that suggest its validity. Two of
them are the framing independence of $\rho_1 (C)$ and $\rho_2(C)$
separately. The framing independence of these objects has been rigorously
proven [\gmm,\dror].  As argued in the previous section, from a quantum field
theory point of view one would expect that the ambiguity present in $n$-point
functions at coincident points would play a role when all $n$ points get
together. Since a Wilson line consists of a path-ordered integration, such
coincident points may occur only for the case of two-point functions
($n=2$), in particular when they are collapsible. By no  means this
argument provides a proof of the theorem but it makes its validity
plausible. In rigorous terms one should think of the theorem above as a
conjecture. In the rest of this section we will find further evidence
regarding its validity. Assuming that the theorem holds the following
corollary follows.

{\bf Corollary.} {\sl  If all the contribution to the self-energy comes from
one loop diagrams, then $\bigl< W(C) \bigr> =F(C;N)\, q^{n(N^2-1)/2N}$
where $F(C;N)$ is framing independent but knot dependent, and the exponential
is manifestly framing dependent but knot independent}.

{\sl Proof.} Let us prove this corollary first forgetting
about the shift $k \to k+N$, or in other words, not including loops.
Let us recall that free propagators are the ones with both endpoints on the
knot.  For example, diagram $b$ of Fig. \lacuatro\ does not
contain free propagators while diagrams $c_2$ and $c_3$
of Fig. \lacuatro\ contain two free propagators. In
diagram $c_2$ of Fig. \lacuatro\ these two free
propagators are collapsible.
To prove the corollary we will organize the perturbative
series expansion of the Wilson line in the following way.
First, select all the diagrams which do not contain free
propagators. Let us denote by ${\cal M}$ the set of
these diagrams. The simplest diagram of this set is
the one with no internal line at all, which is the zeroth
order diagram.
Diagram $b$ of Fig. \lacuatro\ is the order $g^4$ diagram
in the set ${\cal M}$. In virtue of the theorem above
the contribution of each of the diagrams in this set is framing
independent.  Now take each of the diagrams of this set and dress it with
free propagators in all possible ways. Certainly, this organization
exhausts the perturbative series. The proof will consist in demonstrating
that the effect of dressing by free propagators  each diagram in
${\cal M}$  is such that the contribution to the
perturbative series expansion factorizes as stated in the
corollary. To be more specific, we will show that the
form of the contribution of a diagram in ${\cal M}$
plus all the diagrams resulting of its dressing by free
propagators factorizes in a part containing  all the framing dependence
which has the form $q^{n(N^2-1)/2N}$ times a part which is framing
independent.

Let us consider a diagram  $ A\in{\cal M}$ and let us
denote by $\{D_A^p\}$ the set of diagrams resulting
after dressing the diagram $A$ with $p$ free propagators.
This set of diagrams in $\{D_A^p\}$ has been schematically drawn in
\FIG\lasiete{A general set of diagrams with $p$ propagators in the knot} Fig.
\lasiete. Given a diagram in $\{D_A^p\}$, one can work out its
group trace and notice that after commuting
appropriately the generators of $SU(N)$ entering into
this trace in such a way that generators with the same
index get together one generates a series of terms,
being the last of them the group factor of the diagram
in $\{D_A^p\}$ with $p$ collapsible propagators.
For example, the representative of $\{D_A^p\}$ shown in Fig. \lasiete\ would
provide a group structure whose last (or leading) term is as the one of the
diagram  pictured in \FIG\laocho{The less crossed diagram of Fig. \lasiete.}
Fig. \laocho. This procedure is the one which we have followed, for
example, in the derivation of the group factors
\ccuatro, \ccinco\ and \cseis. To gain a better
understanding about the types of group factors which
appear we will consider several subsets of $\{D_A^p\}$. At
first sight one could think that diagrams with $p$ free
propagators with the same number of crossings lead to
the same group structure. This is not entirely true. It holds
for diagrams without three-vertices with one crossing of free propagators but
it
is not true in general. Given a diagram in $\{D_A^p\}$ with $c$ crossings one
finds different group factors. For example one can
check explicitly that the group factor of the diagram
in \FIG\lanueve{A framing independent diagram.}
Fig. \lanueve\ is different than the one of diagram
$g_{12}$ of Fig. \laseis\ with one more (collapsible) propagator.
Let us denote by $\{D_A^{p,c,j}\}$ the
set of diagrams in $\{D_A^p\}$ with $c$ crossings and
group factor $j$. Certainly,
$$
\{D_A^{p,c,j}\}  \subset \{D_A^p\}.
\eqn\fiuno
$$

Given a diagram $D_A^{p,c,j}$ one finds after working out the group factor
that it always contains one which corresponds to the power of order $p$ of
the quadratic Casimir, $[(N^2-1)/2N]^p$ times the group factor
corresponding to diagram $A$. In the process one finds other
group structures with lower powers.
Let us concentrate first on the group structure with the
highest power. Certainly, all diagrams in
$D_A^{p,c,j}$ for a fixed value of $p$ contribute to this group structure.
To apply the factorization theorem of Appendix C we need to have as many
diagrams as domains. As shown at the end of Appendix C, if diagram $A$ is
connected the difference between the number of domains and the number of
diagrams comes about because while diagrams with $n_i$ identical subdiagrams
count as one, from the point of view of domains they should count as $n_i!$ to
have the adequate relabelings to be in the hypothesis of the factorization
theorem. Thus, for the case in which $A$ is connected one just has to repeat
 $p!$ times the diagrams and make the adequate relabelings to be in hypothesis
of the factorization theorem. Of course, this implies that one must divide the
result of the theorem by $p!$. For $A$ connected the contribution corresponding
to the group structure $[(N^2-1)/2N]^p$ from all diagrams in
$\{D_A^{p,c,j}\}$  is,
$$
{1\over p!} {n^p\over 2^p} (ig^2)^p \Big( {N^2-1\over 2N}\Big)^p D_A=
{1\over p!}\Big(in{2\pi\over k}{N^2-1\over 2N}\Big)^p D_A,
\eqn\fidos
$$
where  the
factor $2^p$ appears after enlarging the integral of each propagator to the
whol
   e
knot (which provides the factor $n^p$, where $n$ is the winding number \vtres).
Notice that in \fidos\ $D_A$ represents the contribution from diagram $A$,
which
is framing independent, and that we have used \ocho. Notice also that after
summing in $p$ \fidos\ gives the form stated for the Wilson line in the
corollary. However, this is not the only framing dependent contribution. One
certainly has more contributions  with other group structures. Also, one has
to discuss the situation in which $A$ is not connected. We will consider that
situation later.

To the next group structure (next to leading) not all the diagrams in
$D_A^{p,c,j}$ contribute. Indeed, only diagrams with $c>0$ do. There are,
however, diagrams which contribute and are framing dependent so we have to work
out this dependence. For example, diagrams $g_6$ to $g_{11}$ of Fig. \laseis\
ar
   e
framing dependent. One would like to have enough diagrams to be able to use the
factorization theorem of Appendix C and factorize the contribution as the one
from diagram $A$ with one crossing of free propagators, which is framing
independent, times the contribution due to $p-2$ collapsible propagators which
i
   s
framing dependent and proportional to $n^{(p-2)}$. To explain how one must
arrange the perturbative series to extract the effect of the framing we will
consider first in some detail the case corresponding to diagrams $g_6$ to
$g_{11}$ of Fig. \laseis. This is a particular case in which $p=3$ and $A$ is
trivial but it possesses the essential features  in which we will focus our
attention in the general proof.

Diagrams $g_1$ to $g_{15}$ of Fig. \laseis\  contribute to the group structures
worked out in \ctres, \ccuatro, \ccinco\ and \cseis.
All fifteen diagrams $g_1$ to $g_{15}$
shown in Fig. \laseis\ contribute to the group structure
$-(N^2-1)^3/8N^2$. It corresponds to the general case which led to \fidos.
We will describe it in this example for completeness.
The contribution to this group structure can be written as follows:

\vbox{$$
I_3=\oint_{i<j<k<l<m<n}
\bigl[f(ij,kl,mn)+f(ij,km,ln)+f(ij,kn,lm)+f(ik,jl,mn)
$$
\vskip-2.1cm
$$
\eqalign{\hbox{\hskip3.5cm} &+f(ik,jm,ln)
+f(ik,jn,lm)+
f(il,jk,mn)+f(il,jm,kn) \cr &+f(il,jn,km)+f(im,jk,ln)+f(im,jl,kn)
+f(im,jn,kl) \cr &+
f(in,jk,lm)+f(in,jl,km)+f(in,jm,kl) \bigr] \cr}
\eqn\fdsiete
$$}
\noindent  We have use a notation in which $i,j,k,l,m,n$ represent the six
points attached to the knot and $f(ij,kl,mn)$ the corresponding integrand.
As explained, not all possible domains are represented in \fdsiete. One
possesses 90 domains while there are only 15 diagrams. The ratio between these
two numbers is just 3!, the number of possible orderings of the three free
propagators which from the point of view of domains should be different.
Introducing all the relabelings needed to apply the factorization theorem of
Appendix C one must therefore divide by 3!. The
result is,
$$
I_3={1\over6} \oint_{i<j} \oint_{k<l}\oint_{m<n}p(i,j)\,p(k,l)\,p(m,n),
\eqn\fdocho
$$
where $p(i,j)$ represents the integrand corresponding to a free propagators
with its endpoints $i$ and $j$ attached to the knot.
The linking number $n$ appears when
we let the endpoints of each propagator go freely over the knot
and the frame, and multiply by $1/2$ per propagator. This provides
 the additional
factor  $(1/2)^3=1/8$. The result, in agreement with \fidos\ is,
$$
I_3 = {n^3\over 3!2^3}.
\eqn\fnew
$$

The diagrams contributing to the next group structure, are $g_6$ to $g_{15}$ of
Fig. \laseis, and in order to apply the factorization theorem we need to have
15
diagrams since that is the number of domains. Therefore, we have to overcount
some of them. Notice that now one of the subdiagrams is a free propagator while
the other is $c_3$ of Fig. \lacuatro, which is not connected. We can not apply
the simple strategy described at the end of Appendix C. The number of domains
is
different than the number of diagrams because in diagrams like $g_{12}$,
$g_{13}$ and $g_{14}$ one has two possible choices of domains while in diagrams
$g_{15}$ one has three. Thus let us add what we need, \ie, let us consider two
diagrams of types $g_{12}$, $g_{13}$ and $g_{14}$ and make the apropiate
relabelings in one of each pair, and 3 diagrams of type $g_{15}$ and make
relabelings in two of them. Certainly, we must subtract what we have added.
Now  one can not just simply divide by a factor. We need therefore to subtract
once diagrams $g_{12}$, $g_{13}$ and $g_{14}$ and twice diagram $g_{15}$.
These diagrams are framing independent and therefore they do not contribute to
the framing dependent part. In the general proof, at this stage, the diagrams
which one must subtract contribute to a lower power of $n$. Therefore, we may
use the algorithm safely power by power. Using our previous notation, the
contribution to the next group structure, ${1\over 8}[N^2-1]^2$ is,

\vbox{$$
I_1=\oint_{i<j<k<l<m<n}
\bigl[f(ij,km,ln)+f(jk,ln,mi)+f(kl,jn,im)+f(lm,ik,jn)
$$
\vskip-2.1cm
$$
 \eqalign{\hbox{\hskip4cm} & +f(mn,ik,jl)+
f(ni,jl,km)+f(ik,jm,ln)+f(im,jl,kn) \cr &+f(il,jn,km)+f(il,jm,kn) \bigr] \cr &
\hbox{\hskip-2cm}=\biggl(\oint_{i<j<k<l<m<n}
+\oint_{k<i<j<l<m<n}+\oint_{k<l<i<j<m<n}+\oint_{k<l<m<i<j<n} \cr &
\hbox{\hskip-1.5cm}+
\oint_{k<l<m<n<i<j}+\oint_{i<k<l<m<n<j}
+\oint_{i<k<j<l<m<n}+\oint_{i<k<l<m<j<n}\cr
&\hbox{\hskip-1.5cm}+\oint_{k<i<l<m<n<j}
+\oint_{i<k<l<j<m<n}\biggr)f(ij,km,ln),
\cr} \eqn\fdnueve  $$}
\noindent which, after adding single replicas for the integrands corresponding
t
   o
$g_{12}$, $g_{13}$ and $g_{14}$ and double ones for $g_{15}$, using statement 1
of Appendix C, and making the relabelings,
$$
\sigma_7 = \left( \matrix{1&2&3&4&5&6 \cr 3&4&5&1&6&2
\cr}\right) \quad \hbox{\hskip-0.25cm}\sigma_8 = \left(
\matrix{1&2&3&4&5&6 \cr 3&1&4&2&5&6 \cr}\right) \quad
\hbox{\hskip-0.25cm}\sigma_9 = \left( \matrix{1&2&3&4&5&6 \cr 3&4&1&5&2&6
\cr}\right)  $$
$$
\sigma_{10} = \left( \matrix{1&2&3&4&5&6 \cr 3&1&4&5&2&6
\cr}\right) \quad \sigma_{10}' = \left( \matrix{1&2&3&4&5&6 \cr
3&4&1&5&6&2 \cr}\right)
$$
one finds,
$$
I_1=\tilde I_1 - \hat I_1,
\eqn\renueva
$$
where,
$$
\eqalign{\tilde I_1 =\biggl(&\oint_{i<j<k<l<m<n}
+\oint_{i<k<j<l<m<n}+\oint_{i<k<l<j<m<n}+\oint_{i<k<l<m<j<n} \cr &
+\oint_{i<k<l<m<n<j}+\oint_{k<i<j<l<m<n}
+\oint_{k<i<l<j<m<n}+\oint_{k<i<l<m<j<n}\cr &
+\oint_{k<i<l<m<n<j}
+\oint_{k<l<i<j<m<n}+ \oint_{k<l<i<m<j<n}+\oint_{k<l<i<m<n<j}
\cr & +\oint_{k<l<m<i<j<n}+\oint_{k<l<m<i<n<j}+
\oint_{k<l<m<n<i<j}\biggr)f(ij,km,ln) \cr &
 =\oint_{i<j}p(i,j)
\oint_{k<l<m<n}f_2 (kl,mn).\cr} \eqn\fvdos
$$
and,
$$
\eqalign{
\hat I_1 =\biggl(&\oint_{k<l<i<m<j<n}+\oint_{k<l<i<m<n<j}
 +\oint_{k<l<m<i<j<n}\cr & +\oint_{k<l<m<i<n<j}+
\oint_{k<l<m<n<i<j}\biggr)f(ij,km,ln).\cr}
\eqn\fvdosbis
$$
In \fvdos, the quantity
$f_2$ is the integrand corresponding to diagram $c_3$ of Fig. \lacuatro.
After integration, it results the quantity $\rho_2$ in \tuno.
The contributions remaining to the next group structures plus the left over
represented by \fvdosbis\ are framing independent. This means that we have
extracted all the framing dependence from the set of diagrams $g_1$ to $g_{15}$
of Fig. \laseis. Using \fnew\ and \fvdos\ we can write such a contribution as
$$
\eqalign{
-{1\over 8N^2}&(N^2-1)^3{n^3\over 3! 2^3} + {1\over 8} (N^2-1)^2{n\over
2}\cr &=-{N\over 8}{1\over 3!}\Big({(N^2-1)n\over 2N}\Big)^3
+{N\over 8}(N^2-1){1\over 1!}\Big({N^2-1\over 2N}\Big)^1 \rho_2 \cr}
\eqn\fvtresbis
$$
In the second line of this equation we have rewritten the contribution to show
the general structure which we will find out now.

Let us discuss which one is the algorithm to arrange the perturbative series
expansion to extract the framing dependence for the group structure next to the
leading one.
In the example considered we have seen that diagrams contributing to this
type of group structure were $g_6$ to $g_{15}$ of Fig. \laseis. Certainly,
these
were not all the 15 diagrams which were needed to use the factorization theorem
and factorize the contribution to the group factor  $(N^2-1)$   as a
product of a diagram like the one in Fig. \latres\ times the one originating
$\rho_2$ (diagram $c_3$ of Fig. \lacuatro). However, we were able to  add
pieces of diagrams with the adequate group structure with two or more crossings
(which one must subtract when considering the situation leading to a
factorization of diagrams with two crossings) to have the 15 needed and apply
the factorization theorem of Appendix C. Clearly, this is the procedure which
one must carry out when considering the next to the leading group structure.
One adds pieces of diagrams to complete the set in such a way that the
factorization theorem of Appendix C can be applied, and, on the other hand one
subtracts them. The important point is that all diagrams which are involved in
this operation contain a lower power of $n$, the linking number, and therefore,
in the process one extracts all the framing dependence for the next to leading
group structure.
The corresponding contribution  from all the diagrams in $D_A^{p,c,j}$
for a fixed value of $p$ is,
$$ {1\over (p-2)!} {n^{(p-2)}\over 2^{(p-2)}}
(ig^2)^{(p-2)} \Big( {N^2-1\over 2N}\Big)^{(p-2)} D_A^{(1,i)}= {1\over
(p-2)!}\Big(in{2\pi\over k}{N^2-1\over 2N}\Big)^{(p -2)}D_A^{(1,i)},
\eqn\fitres
$$
where the origin of each factor is similar to the case of \fidos\ and
$D_A^{(1,i)}$ is one of the possible types of diagrams resulting after
dressing diagram $A$ with two free propagators with one crossing.
In \FIG\ladiez{Factorization of $\rho_2$} Fig. \ladiez\ a particular situation
of \fitres\ has been depicted.

One has now to analyze the next group structure. In this case one has to take
into consideration all the left-overs from the previous one. Certainly, this is
going to change the numerical factor in front of this contribution but once
this is taken into account one may proceed similarly as the previous case. It
is
clear now that one may proceed performing this construction for a fixed value
of
$p$ with all types which appear at each value of $c$. The result that one
obtains in this way is,
$$ {1\over p!}\Big(in{2\pi\over k}{N^2-1\over 2N}\Big)^p
D_A+ \sum_{c=1}^{p\choose 2}\sum_{i=1}^{n_c}
 {1\over (p-c-1)!}
\Big(in{2\pi\over k}{N^2-1\over 2N}\Big)^{(p
-c-1)}D_A^{(c,i)},
\eqn\ficuatro
$$
where $n_c$ is the number of types which one finds at $c$ crossings, and
we have used the fact that for the case of $p$ free propagators the
maximum number of possible crossings is ${p\choose 2}$. It is important
to remark that the coefficients $D_A,D_A^{(c,i)}$, which appear in \ficuatro\
are framing independent. In \ficuatro\ we have singled out all the framing
dependence at a given order of free propagators for a diagram $A\in{\cal
M}$. Summing over $p$ one obtains the anticipated exponential behavior:
$$
(D_A+\sum_{c=1}^{\infty}\sum_{i=1}^{n_c}
 D_A^{(c,i)}) \exp\Big({1\over 2}ing^2{N^2-1\over 2N}\Big)=
(D_A+\sum_{c=1}^{\infty}\sum_{i=1}^{n_c}
 D_A^{(c,i)}) q^{n{N^2-1\over 2N}},
\eqn\ficinco
$$
where we have used \ocho\ and \dos. Therefore the corollary is proven.

As this exponential is a common factor,
the sum of the perturbative series factorizes as the sum of all diagrams
without collapsible  propagators (framing independent, but knot
dependent) multiplied by some adequate coefficients times the preceding
exponential (manifestly framing dependent). Note that this factor does not
depend on the kind of knot because the Gauss integral only sees the linking
number. Also is interesting to note that the framing independent and knot
dependent factor is intrinsic to the knot and then includes all the diagrams
that give the building blocks of its topological invariants. This is
schematically represented in \FIG\laonce{Factorization of the framing
dependence.} Fig. \laonce.

All along our discussion regarding the proof of the corollary we have assumed
that $A$ was a connected diagram. Let us remove now that fact. If $A$ is not
connected it is clear that one can use the technic of adding and removing
pieces
of diagrams to apply the factorization theorem at each stage of the proof
described for the case in which $A$ was connected. This will introduce some
numerical factors in \ficinco\ which are important in what regards the building
blocks of the knot, but are irrelevant for the framing dependence since the
exponential behavior has been shown for each term.

Finally we have to include the shift $k \to k+N$. Assuming that  diagrams
with more than one loop do not contribute to the self-energy, we have to
factorize propagators with and without self-energy insertions, and then
sum the resulting series. Remember again that the rest of the diagrams add up
to
    a
framing independent and knot dependent factor.

This series is hard to manage because there can be any number of self-energy
insertions at each propagator, and any number of propagators. The best
organization is as follows: call $\{D_p^q\}$ the set of diagrams with $p$
propagators in which we have inserted $q$ self-energies in all the possible
ways, and $f(n)$, the sum of  all of them, which is the framing dependent and
knot independent factor in $\bigl<W(C)\bigr>$. Notice that,
$$
f(n)=\sum_{p=0}^\infty \sum_{q=0}^\infty \{D_p^q\}.
\eqn\cnueve
$$
The important point to use here is that the distribution of
self-energies is such that they are indistinguishable. The number of
possible distributions of $q$ identical insertions in $p$ lines is a
Bose-Einstein combinatory factor. The lines in which we insert  self-energies
are also indistinguishable. For example,  insertions  done in  sets
of diagrams as the ones in  Fig. \lasiete\  introduces a factor $1/p!$.
Therefore, the prefactor of  $\{D_p^q\}$ is   $$
{1\over p!}{p+q-1\choose q}. \eqn\cdiez
$$
Now, each insertion amounts to a factor $-N/k$  ($k=4\pi /g^2$), and each line
to a factor $x/k$  ($x=in2\pi(N^2-1)/2N$). Hence,
$$
f(n)=\sum_{p=0}^\infty \sum_{q=0}^\infty {1 \over p!}{p+q-1\choose
q}\biggl({x \over k}\biggr)^p \biggl({-N \over k}\biggr)^q.
\eqn\conce
$$
This series in $q$ is the expansion of a factor that provides the shift:
$$
\biggl(1+{N \over k}\biggr)^{-p}=\sum_{q=0}^\infty {p+q-1\choose
q}\biggl({-N \over k}\biggr)^q,
\eqn\cdoce
$$
and therefore, the final result is
$$
f(n)=\sum_{p=0}^\infty {1 \over p!}\biggl({x \over k}\biggr)^p
\biggl(1+{N \over k}\biggr)^{-p}=e^{{in2\pi (N^2-1)\over 2N}{1 \over
{k+N}}}=q^{n{N^2-1\over 2N}},
\eqn\ctrece
$$
where we have used \dos.
This proof also works in the other way around. Suppose that
two-loop diagrams also contribute. These are identical among
themselves, but distinguishable from the one loop insertions, and so there
must appear two ``bosonic'' combinatory factors. The sum of this series
has to be different from the shifted exponential, because expanding the
exponential with shift we find just one bosonic combinatory factor. Then, the
exact result implies that the only quantum corrections relevant to the framing
dependent part are the one-loop self-energies.

This corollary shows from a perturbation theory point of view that all
dependence on the framing in the vacuum expectation value of a knot
factorizes in the form predicted by Witten [\witCS]. Notice also that
assuming that there are only  one-loop quantum corrections, we have found for
an arbitrary knot in the fundamental representation of $SU(N)$ the shift in the
framing dependent factor of the Wilson Line through a purely perturbative
approach.

We have proved that the effect of one-loop contributions on the framing
dependent factorized part of the vacuum expectation value is just a shift in
$k$. Certainly, this is going to hold also for the rest. This, together with
the
factorization of the framing dependence means that we can write from \uno\ the
full contribution from the building blocks or diagrams in ${\cal M}$ with no
loop insertions. One just has to set $n=0$ in \uno\ and remove the shift.

\endpage

\chapter{Conclusion and final remarks}

In this paper we have shown that agreement between the exact result found
by Witten [\witCS] for the vacuum expectation value of the unknot and the
Chern-Simons perturbative series expansion implies that the two-loop
contribution to the one-particle irreducible two-point function must vanishes.
We have worked within a renormalization scheme which is gauge invariant and
which provides a one loop correction to the two and three-point functions
which,
as shown here, is responsible for the shift of $k$ into $k+N$ observed in
[\witCS]. Consistency with the exact result implies that the two-loop
contribution in renormalization schemes providing quantum corrections at one
loop must vanish. Work is under completion regarding this issue [\aalr] for the
renormalization scheme proposed in sect. 2 of this paper.

Our analysis of the structure of the series expansion which appears in the
perturbative calculation of the vacuum expectation value of Wilson lines shows
that in general one can disentangle the framing dependence from the rest of the
contribution. We have shown that under the assumption that the theorem stated
in
sect. 4 holds, the framing dependence of the vacuum expectation value of Wilson
lines factorizes in the form predicted in [\witCS]. We have shown this for a
Wilson line carrying the fundamental representation of $SU(N)$ but it is clear
from the proof that similar arguments hold for any other representation.
Although a rigorous proof of the theorem (conjecture) stated in sect. 4 would
be
very valuable to make our discussion on the factorization of the framing
dependence complete, the arguments based on physical grounds utilized in sect.
3
make the validity of this conjecture rather plausible.

It is important to remark
that the factorization theorem proved in Appendix C has played an essential
role
in the factorization of the framing dependence achieved in sect. 4, as well as
in the explicit calculation of the vacuum expectation value of the unknot in
the
fundamental representation at order $g^6$. In general, this theorem decreases
th
   e
number of integrations needed at a given order by one, reducing the computation
to just the building blocks of the perturbative expansion.

The study carried out in sect. 4 to extract the framing dependence out of the
perturbative series expansions of the vacuum expectation value of a Wilson line
has also provided information about the building blocks of the perturbative
series expansion. Presumably, these building blocks generate a whole series of
topological invariants whose integral form is easy to write down using the
Feynman rules of the theory. Certainly, these building blocks are framing
independent since, according to the corollary of sect. 4, all framing
dependence
has been factorized out. To prove their topological invariance is much harder
and it may well happen that at a given order in $g$ not all the building blocks
by themselves are topological invariants but adequate combinations of them. It
would be desirable to have some general result in this respect.

In this paper we have carried out an explicit calculation of the vacuum
expectation value of the unknot in the fundamental representation of $SU(N)$ up
to order $g^6$. It is straightforward to generalize this calculation  to any
other representation. One would like, however, to analyze the case of a
non-trivial knot to verify if the same conclusion holds and to compute some of
its building blocks.  Chern-Simons theory provides a whole series of
topological
invariants whose integral form is simple (but tedious) to write down, which
would be interesting to classify and characterize. For example, one would like
to know if the degree of complexity of a knot is related to the number of
building blocks which are different from zero. For the case of the unknot we
have that the lemma of sect. 3 plus the theorem of sect. 4
imply that its building blocks are diagrams which do not contain free
propagators. The quantity $\rho_1$ defined in \vocho\ is the first non-trivial
building block. It is represented by diagram $b$ of Fig. \lacuatro. The
building
blocks of the unknot at next order are represented by diagrams $e_1$, $e_2$ and
$e_3$ of Fig. \laseis. As shown in sect. 3 together with Appendix B the
contribution from these diagrams vanishes. Therefore, the next possibly
non-vanishing building blocks for the unknot corresponds to diagrams containing
two three-vertices with all their legs attached to the Wilson line (a
representative is diagram $a$ of \FIG\ladoce{Diagrams at order $g^8$
corresponding to building blocks} Fig. \ladoce) and diagrams containing three
three-vertices (a representative is diagram $b$ of Fig. \ladoce). The
contribution from these building blocks should be computed and compared to the
exact result. As argued at the end of Appendix B, all building blocks of the
unknot corresponding to connected tree-level diagrams with an even number of
vertices vanish. This is in agreement with the full result \uno. From \uno, as
explained at the end of sect. 4, it is rather simple to obtain the contribution
from the building blocks. One has just to set $n=0$ and remove the shift. The
remaining series is clearly even in $1/k$ which implies that only terms at
order
$g^{4m}$ are different from zero, in agreement with the observation made at the
end of Appendix B.

In this work we have shown how to extract from the perturbative series
expansion
of knots in Chern-Simons theory their framing dependence as
well as the effect of quantum corrections. This leaves the series with
the essential ingredients which we have called building blocks and contain all
t
   he
topological information. Further work is needed to study the general
features and the classification of these building blocks.

\endpage

\Appendix{A}

In this appendix we present a summary of our group-theoretical
conventions. We choose the generators of $SU(N)$, $T^a$, $a=1,...,N^2-1$,
to be antihermitian such that
$$
[T^a,T^b] = - f^{abc}T^c,
\eqn\auno
$$
and $f^{abc}$ are completely antisymmetric, satisfying,
$$
f^{acd}f^{bcd} = N \delta^{ab}.
\eqn\aunop
$$
The convention chosen in \auno\ seems unusual but it is the right one when
 the Wilson line is defined as in \dinueve. If we
 had chosen $if^{abc}$ instead of $-f^{abc}$, the exponential of the
Wilson line would have had $ig$ instead of $g$. Our convention also introduces
a
$-1$ in the vertex (see Fig. \launo). The fundamental representation of $SU(N)$
is normalized in such a way that, $$ \tr(T^a T^b) = -{1\over 2} \delta^{ab}.
\eqn\ados $$ The quadratic Casimir in the fundamental representation has the
form, $$
\sum_{a=1}^{N^2-1} T^a T^a = - {N^2-1\over 2N}.
\eqn\atres
$$
One of the group factors which appear in subsect. 3.3 of the paper is
the following,
$$
\tr(T^aT^bT^cT^d) f^{ace}f^{ebd},
\eqn\acuatro
$$
which can be shown to be zero. In fact, using the invariance of the trace
under cyclic permutations one finds, after relabeling,
$$
\tr(T^aT^bT^cT^d) f^{ace}f^{ebd}=
\tr(T^bT^cT^dT^a) f^{ace}f^{ebd}=
\tr(T^aT^bT^cT^d) f^{dbe}f^{eac},
\eqn\acinco
$$
which is just the same as \acuatro\ but with the opposite sign. Therefore,
$$
\tr(T^aT^bT^cT^d) f^{ace}f^{ebd}=0.
\eqn\aseis
$$

\endpage

\Appendix{B}

In this appendix we show that the contribution from diagrams $e_1$ or
$e_2$ of Fig. \laseis\ vanishes for the case of the unknot.
These are the integrals that appear in the four-point $g^6$
contribution to the unknot, represented in diagrams $e_1, e_2$ and $e_3$
of Fig. \laseis.
The idea of the calculation is as follows. According to the framing
independence theorem of sect. 4, each diagram is framing independent, so
we can think that the four points are all in the unknot. Also we assume
that it corresponds to a  topological invariant and therefore we choose the
unknot to be a circumference on the $x_0=0$ plane, centered at the origin.
Call $p$ and $q$ the points of integration over $R^3 \otimes R^3$.
Now observe that the integrand contains an odd number of $\epsilon_{\alpha
\beta \gamma}$ contracted in such a way that it is a pseudoscalar. Its sign
is different in the $x_0>0$ and $x_0<0$ regions. In other words, for each
$(p,q) \in R^3 \otimes R^3$ there are $(p',q') \in R^3 \otimes R^3$ such
that $p_0'=-p_0$, $q_0'=-q_0$ and all other components unchanged, for
which the integrands are equal in magnitude but  different in sign. Then,
in the $p_0, q_0$ plane we have and odd integrand and so the integral
vanishes. Let us verify this explicitly.

The integrations entering this contribution are of the type,
$$
\eqalign{\int d^3p \,d^3q\,&\epsilon_{\mu \rho_1 \nu_1} dx^\mu
{(x-p)^{\rho_1} \over |x-p|^3}\epsilon_{\nu \rho_2 \nu_2} dy^\nu
{(y-p)^{\rho_2} \over |y-p|^3}\epsilon^{\nu_1 \nu_2
\tau_1}\epsilon_{\rho \rho_3 \nu_3} dz^\rho {(z-q)^{\rho_3} \over
|z-q|^3} \cr &\epsilon_{\tau \rho_4 \nu_4} dw^\tau
{(w-q)^{\rho_4} \over |w-q|^3}\epsilon^{\nu_3 \nu_4
\tau_2} \epsilon_{\tau_1 \tau_2
\rho_5}{(p-q)^{\rho_5} \over |p-q|^3},\cr}
\eqn\buno
$$
where $x,y,z,w$ lie on the knot and hence $x_0=y_0=z_0=w_0=0$.
The denominator of the integrand of this expression is
invariant under $p_0, q_0 \to -p_0, -q_0$. The
structure of the numerator is  more complicated and we will consider its
form separately. The first three factors of the numerator of \buno\ become,
 $$
\eqalign{
&\epsilon_{1\rho_1 \nu_1} dx^1 (x-p)^{\rho_1} +
\epsilon_{2\rho_1 \nu_1} dx^2 (x-p)^{\rho_1} \cr
& =  \epsilon_{1 0 \nu_1} dx^1 (x-p)^0 +
\epsilon_{1 2 \nu_1} dx^1 (x-p)^2 +
\epsilon_{2 0 \nu_1} dx^2 (x-p)^0 +
\epsilon_{2 1 \nu_1} dx^2 (x-p)^1. \cr}
\eqn\btres
$$
The following three factors give an analogous contribution,
$$
\eqalign{
&\epsilon_{1\rho_1 \nu_2} dy^1 (y-p)^{\rho_1} +
\epsilon_{2\rho_1 \nu_2} dy^2 (y-p)^{\rho_1} \cr
&=  \epsilon_{1 0 \nu_2} dy^1 (y-p)^0 +
\epsilon_{1 2 \nu_2} dy^1 (y-p)^2 +
\epsilon_{2 0 \nu_2} dy^2 (y-p)^0 +
\epsilon_{2 1 \nu_2} dy^2 (y-p)^1, \cr}
\eqn\bcuatro
$$
which becomes \btres\ after changing $y \to x$ and $\nu_2 \to \nu_1$.
Contracting \btres\ and \bcuatro\  with $\epsilon^{\nu_1 \nu_2 \tau}$
one obtains,
$$
\eqalign{&\epsilon_{2 0 \tau_1} \bigl[ dx^1 p^0 dy^1 (y-p)^2
-dx^1 p^0 dy^2 (y-p)^1 - dx^1 (x-p)^2 dy^1 p^0 +
dx^2 (x-p)^1 dy^1 p^0 \bigr] \cr
-&\epsilon_{0 1 \tau_1} \bigl[ dx^1 p^0 dy^2 (x-p)^2
-dx^2 p^0 dy^1 (y-p)^2 + dx^2 (y-p)^1 dy^2 p^0 -
dx^2 (x-p)^1 dy^2 p^0 \bigr] \cr +&
\epsilon_{2 1 \tau_1} \bigl[ -dx^1 (p^0)^2 dy^2 +dx^2 (p^0)^2 dy^1
\bigr]. \cr }
\eqn\bcinco
$$
The rest of the factors in \buno\ except the last two are treated
similarly, obtaining an  expression similar to \bcinco\  with $x \to z$, $y
\to w$, $p \to q$, and $\tau_1 \to \tau_2$. Finally, multiplying \bcinco\ by
the remaining factor of \buno,
$\epsilon_{\tau_1 \tau_2 \rho_5}(p-q)_{\rho_5}$,one gets,
$$
\eqalign{
& -p_0 q_0^2 (p-q)_2 \bigl[\cdots \bigr] +p_0 q_0 (p-q)_0
\bigl[\cdots \bigr] -q_0 p_0^2 (p-q)_2 \bigl[\cdots \bigr] -
q_0 p_0^2 (p-q)_1 \bigl[\cdots \bigr] \cr
& +p_0 q_0 (p-q)_0
\bigl[\cdots \bigr] -p_0 q_0^2 (p-q)_1 \bigl[\cdots \bigr], \cr}
\eqn\bseis
$$
where by $\bigl[\cdots \bigr]$ it is meant a part that does not depend
on $p_0$ or $q_0$. As argued above, this expression is odd under
$p_0, q_0
\to  -p_0, -q_0$ and therefore the integration over $p_0, q_0$
in \buno\ vanishes. This way of showing the vanishing of integrations as
\buno\ suggests that this property is a general feature of tree level
connected diagrams with an even (and non zero) number of $R^3$ points of
integration. As we discuss in sect. 5, this assertion is substantiated by the
full result \uno.

\endpage

\Appendix{C}

In this appendix we state and prove the factorization theorem.
First, let us introduce some notation.
We will be considering  diagrams corresponding to a given order
$g^{2m}$ in the perturbative expansion of a knot, and to a given number of
points running over it, namely $n$. Note that  $n$ and $m$ fix the
number of vertices, $n_v$, and propagators, $n_p$,
that are in each diagram: $n_v = 2m-n$
and $n_p =3m-n$.
We will denote by $\{i_1,i_2,\ldots,i_n\}$ a domain of integration
where the order of integration is $i_1<i_2<...<i_n$, being
$i_1,i_2,\dots,i_n$ the points on the knot (notice the condensed
notation) where the internal lines of the  diagram are attached. The
integrand corresponding to that diagram will be denoted as
$f(i_1,i_2,\ldots,i_n)$. Diagrams are  in general composed of
subdiagrams, which may be connected or non-connected. For a given
diagram we will make specific choices of subdiagrams depending on
the type of factorization which is intended to achieve. For example,
for a diagram like $g_6$ of Fig. \laseis\ one may choose as
subdiagrams the three free propagators, or one may choose a
subdiagram to be the collapsible  propagator and other subdiagram to
be the one built by two crossed free propagators.
We will consider a set of diagrams ${\cal N}$ corresponding to a
given order $g^{2m}$, to given number of points attached to the knot,
$n$, and to a given kind. By kind we mean all diagrams containing
$n_i$  subdiagrams of type $i$, $i=1,...,T$. By $p_i$ we will denote
the number of points which a subdiagram of type $i$ has attached to the
knot. For example, if one considers diagrams at order $g^6$ with $n=6$
points attached to the knot, with three subdiagrams which are just
free propagators, this set is made out of diagrams $g_1$ to $g_{15}$ of
Fig. \laseis. However, if one considers diagrams at order $g^6$ with
$n=6$ with a subdiagram consisting of a free propagator and another
subdiagram of the type $c_3$ of Fig. \lacuatro, this set is made out
of diagrams $g_6$ to $g_{15}$.  The contribution from all diagrams in
${\cal N}$ can be written as the following sum:
$$
\sum_{\sigma\in \Pi_n} \oint_{i_1,i_2,\ldots,i_n}
f(i_{\sigma(1)},i_{\sigma^(2)},\ldots,i_{\sigma(n)})
\eqn\newuno
$$
where $\sigma\in \Pi_n$, being $\Pi_n\subset P_n$ a subset of the
symmetric group of $n$ elements. Notice that $\Pi_n$ reflects the
different shapes of the diagrams in ${\cal N}$. In \newuno\ the
integration region has been left fixed for all the diagrams and one has
introduced  different integrands. One could have taken the
opposite choice, namely, one could have left fixed the integrand and
sum over the different domains associated to ${\cal N}$. The first
statement regarding the factorization theorem just refers to these
two possible choices. Let us define the domain resulting of permuting
$\{i_1,i_2,\ldots,i_n\}$ by an element $\sigma$ of the symmetric group
$P_n$ by
$$
d_\sigma=\{\,i_{\sigma(1)},i_{\sigma(2)},\ldots,i_{\sigma(n)}\,\},
\eqn\funo
$$
then the following result immediately follows.

{\bf Statement 1.} {\sl The contribution to the
Wilson line of the sum of diagrams whose integrands are of the form:
$$
f(i_{\sigma (1)}, i_{\sigma(2)}, \dots, i_{\sigma(n)}),
\eqn\ftres
$$
where $\sigma$ runs over a
given subset $\Pi_n \in P_n$ with a common domain of integration is
equal to the sum of the integral of $f(i_1,i_2,\ldots,i_n)$ over
$d_\sigma$ where $\sigma \in \Pi_n^{-1}$:
$$
\oint_{i_1,i_2,\ldots,i_n} \sum_{\sigma \in \Pi_n}
f(i_{\sigma(1)},i_{\sigma^(2)},\ldots,i_{\sigma(n)}) =
\sum_{\sigma \in \Pi_n^{-1}} \oint_{d_\sigma}
f(i_1,i_2,\ldots,i_n).
\eqn\fcuatro
$$}

The idea behind the factorization theorem is to organize the diagrams
in ${\cal N}$ in such a way that one is summing over all
possible permutations of domains. Summing over all domains implies that
one can consider the integration over the points corresponding to each
subdiagram as independent and therefore one can factorize the
contribution into a product given by the integrations of each
subdiagram independently. Our aim in the rest of this appendix will
be to prove the following statement.

{\bf Statement 2.} (Factorization theorem) {\sl Let $\Pi_n '$ be the
set of all possible permutations of the domains of integration of
diagrams containing subdiagrams of types $i=1,...,T.$
If $\, \Pi_n ^{-1} = \Pi_n '$, the sum
of integrals over $d_\sigma$, $\sigma \in \Pi_n^{-1}$, is the product of the
integrals of the subdiagrams over the knot,
being the domains all independent,
$$
\sum_{\sigma \in\Pi_n^{-1}}
\oint_{d_\sigma} f(i_1,\ldots,i_n) =
\prod_{i=1}^{T} \biggl(\oint_{i_1,\ldots,i_{p_i}}
 f_i(i_1,\ldots,i_{p_i}) \biggr)^{n_i},
\eqn\fcinco
$$
In \fcinco\ $n_i$ denotes the number of subdiagram of
type $i$ and $p_i$ its number of points attached to the knot. }

The proof of this statement is  trivial since having all possible
domains it is clear that one can write the integration considering
subdiagram by subdiagram, the result being the product of all the
partial integrations over subdiagrams.

In sect. 4 we considered situations where we were forced to add and
subtract pieces of diagrams in such a way that the theorem above was
utilized. For completeness, we will show now that for the case in
which all subdiagrams are connected the overcounting needed to apply
the theorem is very simple and that it just amount to divide by an
adequate combinatory factor. Let us discuss first an example to
understand the strategy leading to the general situation.

Let us consider the four point $g^4$ contribution or, better to
say, its part with $(N^2-1)^2/{4N}$ as $SU(N)$ factor. The
diagrams are $c_1, c_2$ and $c_3$ of Fig. \lacuatro,
whose contribution can be
written according to Statement 1, is,
$$
\eqalign{
\oint_{i_1 < \ldots <i_4} \bigl[ & f(i_1,i_2,i_3,i_4) +
f(i_1,i_3,i_2,i_4) + f(i_1,i_4,i_2,i_3)\bigr]\cr
&=\biggl(\oint_{i_1< i_2<i_3<i_4}
+\oint_{i_1<i_3<i_2<i_4}+\oint_{i_1<i_3<i_4<i_2}\biggr)
p(i_1,i_2) \, p(i_3,i_4),\cr}
\eqn\fsiete
$$
where $p(i_1,i_2)$ represents a free propagator attached to the knot
at points $i_1$ and $i_2$.  There are no more than three
diagrams, and the number of domains is six. These are:
$$
\eqalign{
&i_1<i_2<i_3<i_4 \cr
&i_1<i_3<i_2<i_4 \cr
&i_1<i_3<i_4<i_2 \cr}
\qquad\qquad
\eqalign{
&i_3<i_1<i_2<i_4 \cr
&i_3<i_1<i_4<i_2 \cr
&i_3<i_4<i_1<i_2 \cr}
\eqn\focho
$$
so we need three more diagrams. Now the
overcounting consists of rewriting the diagrams after a relabeling. In
the course of the relabeling we will use the fact that
$p(i_1,i_2)=p(i_2,i_1)$.
The relabelings needed in the overcounting are the following:
$$
\sigma_1 = \left( \matrix{1&2&3&4 \cr 3&4&1&2
\cr}\right) \quad \sigma_2 = \left( \matrix{1&2&3&4 \cr 3&1&4&2
\cr}\right) \quad \sigma_3 = \left( \matrix{1&2&3&4 \cr 3&1&2&4
\cr}\right) \eqn\fnueve
$$
Here the subindex of each $\sigma$ indicates the integrand over
which it acts. Note also that each relabeling is in fact a repetition of the
diagram, due to the symmetry of the propagator pointed out above.
Therefore, all we have to do is to multiply the sum of the six
terms by $1/2$. The new integrands are:
$$
\eqalign{
\sigma_1 [f(i_1,i_2,i_3,i_4)] =&
f(i_3,i_4,i_1,i_2)=f(i_1,i_2,i_3,i_4),\cr
 \sigma_2 [f(i_1,i_3,i_2,i_4)] =&
f(i_3,i_4,i_1,i_2)=f(i_1,i_2,i_3,i_4),\cr
 \sigma_3 [f(i_1,i_4,i_2,i_3)] =&
f(i_3,i_4,i_1,i_2)=f(i_1,i_2,i_3,i_4),\cr}
\eqn\fnueve
$$
and the result is:
$$
\eqalign{\oint_{i_1 < \ldots <i_4}& \bigl[
f(i_1,i_2,i_3,i_4) + f(i_1,i_3,i_2,i_4) + f(i_1,i_4,i_2,i_3)\bigr]
\cr &= {1\over 2} \biggl(\oint_{i_1< i_2<i_3<i_4}
 +\!\oint_{i_1<i_3<i_2<i_4}+\!\oint_{i_1<i_3<i_4<i_2} \cr &
\hbox{\hskip0.8cm}+\!
\oint_{i_3<i_1<i_2<i_4}
+\!\oint_{i_3<i_1<i_4<i_2}+\!\oint_{i_3<i_4<i_1<i_2}\biggr)
 p(i_1,i_2)\, p(i_3,i_4). \cr}
\eqn\fdiez
$$
Now, in the language of the theorem,
$\Pi_4^{-1} = \Pi_4'$, \ie, each propagator runs freely over the
knot and therefore we obtain,
$$
\oint_{i_1 < \ldots <i_4} \bigl[
f(i_1,i_2,i_3,i_4) + f(i_1,i_3,i_2,i_4) + f(i_1,i_4,i_2,i_3)\bigr]=
 {1\over 2} \oint_{i_1<i_2}\oint_{i_3<i_4} p(i_1,i_2) \,
p(i_3,i_4).
\eqn\fonce
$$
The linking number of the frame and the knot arises as an integral of
a propagator  with one of its endpoints running over the knot and the
other over the frame, without any ordering. This is easily achieved in
the integrals we have by simply leaving $i_1$ and $i_2$ free, and
multiplying by $1/2$. The same should be done for the pair $i_3$ and
$i_4$. This is again an example of factorization. The final $1/8$
is the factor $3/4!$ that appears for $w_4^{(c)}$ in \tdos.

The example suggests the idea of a general
proof. We should count the number of domains and the number of
diagrams and observe their relation, as well as the origin of their
difference. We are able to provide formulae for the number of domains
for an arbitrary diagram. However, our formula for the number of
diagrams in terms of the features of their subdiagrams only holds
when all subdiagrams are connected. Using these formulae we will show
that we can make equivalent the overcounting and the original
contribution simply introducing a combinatory factor.

First, let us compute the number of domains, $d'$, corresponding to
 a general set of diagrams
at a given order in the perturbative expansion of the
knot, to the same number of points on the knot, and to the same types
of subdiagrams. Suppose that we construct the diagram adding its
sudiagrams in a given order. The first one can put its $p_i$ points in
$n$ places. The second one has to distribute its points in the
remaining $n-p_i$, and so on. For example, if there are only $a$
points attached by propagators and $b$ points attached by
three-vertices (and so
$n=2a+3b$),   $$ \eqalign{
d'=&{n \choose 2}{n-2 \choose 2}{n-4 \choose 2}\ldots{n-2a+2
\choose 2}{n-2a \choose 3}{n-2a-3 \choose 3}\ldots \cr
&\hbox{\hskip4.1cm}\ldots {n-2a-3b+3
\choose 3}
 ={n! \over
(2!)^a(3!)^b}.\cr}
\eqn\fvdosp
$$
In general the denominator will include the product of every
$p_i!$, where $p_i$ denotes the number of points attached to the knot
corresponding to a subdiagram of type $i$. If we denote by $d$ the
number of diagrams of the set under consideration, it is clear that
 $d \le d'$ due to the possible
identity of some subdiagrams. In those cases we have to divide $d'$ by
the number of permutations of all the identical subdiagrams. If some
of the subdiagrams were not connected we would have to consider
additional factors which would imply to introduce more data about
each subdiagram. Therefore, let us restrict ourselves to the case of
connected subdiagrams. The final form of the formulae is:
$$ d'={n!
\over \prod_{i=1}^k (p_i !)^{n_i}},  \qquad d={n! \over \prod_{i=1}^k
n_i ! (p_i !)^{n_i}}. \eqn\fvtres $$
The relation between $d$ and $d'$ is always an integer: $$
{d' \over d}=\prod_{i=1}^k n_i !
\eqn\fvcuatro
$$
This is the combinatory factor we were searching for.
Therefore, when the subdiagrams are connected one just has to
overcount evenly each diagram to have as many diagrams as domains and
divide by \fvcuatro.  For  non-connected subdiagrams the previous
formula for $d$ fails. An example is the factorization of $\rho_2(C)$
in the ${1\over 8}(N^2-1)^2$ part of
diagrams $g_6,\ldots,g_{15}$ of Fig. \laseis. The subdiagram that we
would factorize is the corresponding to  diagram $c_3$ of Fig.
\lacuatro, which is not connected. The previous formula gives
$d=d'=15$, exactly the same result as if the subdiagram were $e_1$ or
$e_2$ of Fig. \laseis, but there are just 10 diagrams. The number of
domains, however, is correct.

\endpage
\figout

\endpage

\refout

\endpage

\end